\documentclass[twocolumn,showpacs,preprintnumbers,amsmath,amssymb]{revtex4}
\usepackage{amssymb}
\usepackage{amsmath}
\usepackage{graphicx}
\usepackage{dcolumn}
\usepackage{bm}
\usepackage{color}
\usepackage{exscale}
\usepackage{relsize}
\usepackage{extarrows}
\usepackage{xcolor}

\begin{document}

\title{Zeno effect of the open quantum system in the presence of 1/f noise }
\author{Shu He$^{1,}$}\email{heshu1987@foxmail.com}
\author{Chen Wang$^{2,}$}\email{wangchenyifang@gmail.com}
\author{Li-Wei Duan$^{3}$}
\author{Qing-Hu Chen$^{3}$}\email{qhchen@zju.edu.cn}
\address{
$^{1}$ Department of Physics and Electronic Engineering, Sichuan Normal University, Chengdu 610066, China \\
$^{2}$ Department of Physics, Hangzhou Dianzi University, Hangzhou, Zhejiang 310018, China \\
$^{3}$ Department of Physics, Zhejiang University, Hangzhou 310027, P. R. China
 }

 \date{\today }

\begin{abstract}
  We study the quantum Zeno effect (QZE) and quantum anti-Zeno effect (QAZE) in a two-level system(TLS) interacting with an environment owning 1/f noise. Using a numerically exact method based on the thermo field dynamics(TFD) theory and the matrix product states(MPS), we obtain exact evolutions of the TLS and bath(environment) under repetitive measurements at both zero and finite temperatures. At zero temperature, we observe a novel transition from a pure QZE in the short time scale to a QZE-QAZE crossover in the long time scale, by considering the measurement induced non-Markvoian effect.
  At finite temperature, we exploit that the thermal fluctuation suppresses the decay of the survival probability in the short time scale, whereas it enhances the decay in the long time scale.
\end{abstract}

\maketitle

\section{Introduction}

The quantum Zeno effect(QZE) and quantum anti-Zeno effect (QAZE), that the dynamical evolution  become slowed down and accelerated by frequently measuring the quantum system respectively,
have been extensively considered as powerful strategies to implement the quantum control including quantum information protection\cite{barenco1997stabilization}, decoherence suppression\cite{beige2000quantum}, purification and cooling\cite{erez2008thermodynamical}.
Early investigations of the QZE in open quantum systems were restricted to some exactly solvable models, under the assumption of rotating wave approximation(RWA) and Markovian approximation, in which results were only reliable with the weak system-bath coupling and short time limit of the bath memory \cite{kofman2000acceleration,facchi2000spontaneous,facchi2001quantum}.
Recently, studies based on the  polaron transformation and perturbative expansion, have already exhibited that the conter-rotating terms (CRTs) play a significant role for both QZE and QAZE\cite{zheng2008quantum,cao2010dynamics}.
Moreover, numerical studies, e.g., the hierarchical equations of motion(HEOM) method, indicate that non-markovianity of the bath may be favorable for the accessibility of the QZE\cite{wu2017quantum}.
However, these approaches often only become available for bath with Ornstein-Uhlenbeck-type correlation function\cite{tanimura1989time,yin2012spin,dijkstra2010non,wang2013exact}.
Therefore, it is of great importance to develop an exact method to analyze the QZE/QAZE in the open quantum system coupled to a dissipative environment
with arbitrary spectrum function.

1/f noise, a representative spectrum case of the bath, describes the environment with a spectral density scales inversely proportional to the frequency\cite{dutta1981low}.
It is encountered in broad areas, such as quantum dots in semiconductor systems\cite{chirolli2008decoherence}, Josephson qubits in Cooper-pair box\cite{nakamura2002charge} doped silicon in electron-spin-resonance transistors\cite{vrijen2000electron} and electrons on liquid helium\cite{platzman1999quantum}.
Typically, 1/f noise is used to describe the quantum fluctuation of a qubit, which is induced by charge fluctuations in various quantum computer implementations, e.g., trapped ions\cite{turchette2000heating} and superconducting qubits\cite{nakamura2002charge}.
However, due to the long time correlations of the 1/f noise, the frequently adopted Markov approximation\cite{breuer2002theory,burkard2009non} in the theory of the open quantum system, breaks down.
Moreover, recently realization of strong and ultrastrong couplings between qubit and its environment\cite{bourassa2009ultrastrong,abdumalikov2008vacuum} can not be described by the rotating wave approximation(RWA).
Hence, an exact description of the QZE of an open quantum system in the presence of the 1/f noise bath will  be appreciated both from theoretical and experimental views.
Besides, one of the assumptions widely used in previous studies of QZE, demonstrates that each measurement projects  the system together with its environment to their initial state.
This assumption is valid when the system-environment coupling is weak or the measure frequency is fast enough such that the environment could hardly evolves between measurements.
Our previous study\cite{he2017zeno} has already showed that such measurement induced non-Markovian effect leads to a qualitative modification of the QZE and QAZE in Ohmic and sub-Ohmic environments.
Hence, due to the long-time correlations  of 1/f noise bath, such measurement-induced disturbance on the environment should not be ignored.
Moreover, at finite temperature, it has been shown that the temperature plays an important role on the crossover from QZE to QAZE\cite{maniscalco2006zeno}as well as the formation of the Zeno subspace\cite{montina2008quantum,de2017dynamics,militello2011quantum}.
Thus, it is natural to raise an interesting question: \emph{how does this measurement-induced environmental disturbance affects the QZE(QAZE) in the 1/f noise bath?
What is the influence of the temperature on the QZE(QAZE) in the 1/f noise bath?}

To address the above-mentioned problems,  we employ a highly efficient and numerically exact method, to study the QZE and QAZE of a two-level system interacting with a 1/f noise bath.
It is based on a recently proposed time-dependent variational principle (TDVP) for matrix product states (MPS)\cite{haegeman2016unifying}.
By releasing the original assumption which ignores the disturbance of measurements on the environment,  we show that repetitive measurements drive both TLS and the bath to a non-equilibrium dynamical steady state.
 By increasing the system-environment coupling strength, we observe a novel transition from a pure QZE to a crossover of QZE-QAZE at zero temperature.
 Moreover, We generalized our study to the finite temperature case  by combining the MPS-TDVP method with the so-called thermo field dynamics method(TFD) which has been used recently in the study of electron-vibrational dynamics \cite{borrelli2016quantum,ritschel2015non}.
 We find that the thermal fluctuations raise an additional suppression of decay of the survival probability in the short time regime, whereas it enhances the decay rate in long time scale.
 However, Such combination of two effects blurs the crossover of QZE-QAZE.

This paper is organized as follows. In Sec \uppercase\expandafter{\romannumeral2}, we briefly introduce the MPS-TDVP method combined with TFD to deal with the finite temperature dynamics for the open quantum system.
We show the validity of the approach by comparing with generally recognized results, which were obtained from the numerically exact HEOM method.
In Sec \uppercase\expandafter{\romannumeral3}, we use previously introduced method to study the quantum Zeno dynamics in a 1/f noise bath at zero temperature by taking consideration of the measurement-induced disturbance. Finally, we generalized our study to the finite temperature case in Sec \uppercase\expandafter{\romannumeral4}. We close this paper with a short summary in Sec \uppercase\expandafter{\romannumeral5}.

\section{Model and Methodology}

\subsection{ Thermofield dynamics theory}
%\begin{align}
% \hat{H} = H - \tilde{H}
%\end{align}
%where $\tilde{H}$ represents a fictitious Hamiltonian which can be determined from the original Hamiltonian $H$\cite{}. Then the dynamics of the $H$ can be descried by the following TFD Schr\"{o}dinger equation:
%\begin{align}
% i\frac{\partial}{\partial t}|\psi(t)\rangle = \hat{H} |\psi(t)\rangle
%\end{align}
%where $|\psi(t)\rangle$ is the TFD wave function defined as
%\begin{align}
% |\psi(t) = \rho(t)^{\frac{1}{2}}|\bm{I}\rangle,  \quad |\bm{I}\rangle =\sum_n|n\rangle|\tilde{n}\rangle \label{TFDini}
%\end{align}
%$\rho(t)$ is the density operator of the original physical system while $|n\rangle $($|\tilde{n}\rangle$) are arbitrary basis vectors of the physical(fictitious) space. The expectation value of any operator $A$ acting on the physical space $|k\rangle$ thus can be obtained by
%\begin{align}
% \langle A(t)\rangle = \langle \psi(t)|A|\psi(t)\rangle =\text{Tr}\left[\rho(t)A\right]
%\end{align}

In this work, we consider a spin-boson model of a qubit(TLS) interacting with the environment described by harmonic oscillators. The Hamiltonian is described as:
\begin{align}
 &H = H_{\text{sys}} +H_{\text{env}} +H_{\text{int}}\notag\\
 &H_{\text{env}} =\sum_k \omega_k a^\dag_k a_k \notag\\
 &H_{\text{int}} = \frac{1}{2}\hat{A} \sum_k g_k(a_k^\dag + a_k)\label{SBH1}
\end{align}
where $H_{\text{sys}}$ is the qubit Hamiltonian and  $a_k^\dag $ ($a_k$) denotes the bosonic creation(annihilation) operator corresponding to the environmental mode with frequency $\omega_k$. $g_k$ is the coupling strength between qubit and each mode in a linear interaction form of the displacements $a^\dag_k +a_k$ multiplying a qubit operator $\hat{A}$. Throughout this paper, we  assume the initial state is a  product state  with environment in its thermal equilibrium:
\begin{align}
 \rho(0) = |\psi_{\text{sys}}\rangle \langle \psi_\text{sys}|\otimes Z^{-1}\exp\left(-\beta H_{\text{env}}\right)\label{Phyini}
\end{align}
where $\beta = 1/(k_B T)$ and $k_B$ is the Boltzmann constant.

According to the thermo-field dynamics(TFD) method introduced first in Ref.\cite{suzuki1985thermo,suzuki1991density}, the time evolution of the Hamiltonian $(\ref{SBH1})$ at finite temperature can be described by a TFD Schr\"{o}dinger equation of a modified Hamiltonian $\hat{H}$ defined in an enlarged Hilbert space with double degree of freedom to the original physical Hilbert space:
\begin{align}
 \hat{H} = H +\tilde{H}
\end{align}
where $\tilde{H}$ is called fictitious Hamiltonian, written as:
\begin{align}
\tilde{H}_{\text{env}} = \sum_k (-\omega_k) b_k^\dag b_k
\end{align}
$b_k^\dag$($b_k$) are creation(annihilation) boson operators defined in the fictitious Hilbert
space with the same (but negative) frequency $\omega_k$ as the physical bath $H_{\text{env}}$. Note that since the qubit is initially uncoupled to the environment, its degree of freedom has already been removed from the fictitious Hilbert space\cite{borrelli2016quantum}. By further applying a temperature-dependent Bogoliubov transformation\cite{barnett1985thermofield,takahashi1996thermo}, one can define a new group of bath annihilation and corresponding creation operators:
\begin{align}
  A_k = \sqrt{\bar{n}_k +1}a_k -\sqrt{\bar{n}_k}b_k^\dag \notag\\
  B_k = \sqrt{\bar{n}_k +1}b_k -\sqrt{\bar{n}_k}a_k^\dag \label{ThermoOp}
\end{align}
where $\bar{n}_k = \left(e^{\beta\omega_k} -1\right)^{-1}$ is the mean thermal occupation number of the physical mode $\omega_k$. The initial TFD wave function corresponding to  ($\ref{Phyini}$) in the transformed representation can be simply written as:
\begin{align}
  |\psi(0)\rangle = |\psi_{\text{sys}} \rangle\otimes|\bm{0}(\beta)\rangle
\end{align}
where $ |\bm{0}(\beta)\rangle \equiv |0\rangle|\tilde{0}\rangle$ is the vacuum thermal state for the physical(and fictitious) bath. The total  Hamiltonian for the TFD evolution after the transformation reads:
\begin{align}
  &\hat{H} = H_{\text{sys}} + \tilde{H}_{\text{env}} + \tilde{H}_{\text{int}} \notag\\
  &\tilde{H}_{\text{env}} = \sum_k\omega_k \big(A^\dag_k A_k - B^\dag_k B_k\big) \notag\\
  &\tilde{H}_{\text{int}} =\frac{1}{2} \hat{A} \sum_k g_k\bigg( \sqrt{n_k +1}( A_k^\dag +A_k) +
  \sqrt{n_k}( B_k^\dag +B_k) \bigg) \label{SBTFD}
\end{align}

Since the Bogoliubov transformation only affects on bath modes, the expectation value of an arbitrary operator of the qubit $O_{\text{sys}}$ can be straightforwardly calculated  under TFD framework:
\begin{align}
 \langle O_{\text{sys}}(t)\rangle = \text{Tr}\bigg[ \langle \psi(t)|O_{\text{sys}} |\psi(t)\rangle \bigg]
\end{align}

\subsection{MPS-TDVP method}

In the following, we use time-dependent matrix product state(tMPS) method to efficiently simulate the wave function dynamics of the Hamiltonian (\ref{SBTFD}). Since MPS-based methods work particularly well on one-dimensional model with short-range interaction, we first transform it into a representation of an one-dimensional semi-infinite chain with nearest interaction. This can be realized by using an orthogonal polynomial mapping for both ($A_k^\dag, A_k$) and $(B_k^\dag, B_k$) bath modes(see details in Ref. \cite{chin2010exact}). Thus finally, we obtain our final working Hamiltonian:
\begin{align}
   \hat{H} =  & H_\text{sys} + H^{R}+ H^{L} \\
    H^{L} = &-\sum_k \epsilon_k^{L} c_k^{\dag} c_k - \sum_{k=0}^N t_k^{L}\big(c_k^\dag c_{k+1} +c_{k+1}^\dag c_k\big)\notag\\ &+\frac{1}{2}\kappa_0^L\hat{A}(c_k^\dag +c_k) \\
   H^{R} = &\sum_k \epsilon_k^{R} d_k^{\dag} d_k + \sum_{k=0}^N t_k^{R}\big(d_k^\dag d_{k+1} +d_{k+1}^\dag d_k\big) \notag\\ &+\frac{1}{2}\kappa_0^R\hat{A}(d_k^\dag +d_k) \label{FinalHami}
\end{align}
where $c^\dag_k, c_k$ and $d^\dag_k, d_k$ are new bosonic operators transformed from $B_k^\dag, B_k$ and $A_k^\dag, A_k$ defined in Eq.(\ref{ThermoOp}). $\epsilon_k^{L,R}, t_k^{L,R} $ and $\kappa_0^{L,R}$ can be determined with well-defined procedure\cite{chin2010exact} by corresponding temperature renormalized spectral functions $J_{L,R}(\omega) $:
\begin{align}
 &J_{L}(\omega) = \sum_k g_k^2\bar{n}_k\delta(\omega-\omega_k) = J(\omega)\big(e^{\beta\omega} -1\big)^{-1} \notag\\
 &J_{R}(\omega) = \sum_k g_k^2(\bar{n}_k+1)\delta(\omega-\omega_k)
 = J(\omega)\big(1- e^{-\beta\omega}\big)^{-1} \label{JwTemp}
\end{align}
where $J(\omega) =\sum_k g_k^2\delta(\omega - \omega_k)$ is the spectral function of the physical environment for the original Hamiltonian ($\ref{SBH1}$).
It has been confirmed that \cite{schollwock2011density,schroder2016simulating} MPS-based methods are highly efficient and reliable to simulate the dynamics of such Hamiltonian  which demonstrates an one-dimensional chain structure with nearest interactions. Moreover, instead of using the traditional evolution method of MPS, we employ the recently proposed  time-dependent variational principle(TDVP) method\cite{haegeman2016unifying} to simulate the evolution. This method derives an  optimal equation of motion for each site of the one-dimensional chain by projecting the the Schr\"{o}dinger  equation onto the tangent space of the MPS manifold  based on the Dirac-Frenkel variational principle. On the one hand, it shares a similar form with the original MPS algorithm based on the the Suzuki-Trotter decomposition\cite{suzuki1990fractal} which evolves the wave function site by site with small time steps $\delta t$. On the other hand, TDVP avoids direct decomposition of the propagating operator $U(t)= e^{-iH_{\text{chain}}\delta t}$ thus the error raises only in the integration scheme. Recently,  TDVP has been used to capture the essential long time dynamics of thermalizing quantum systems\cite{leviatan2017quantum} and quantum spin systems
with long-range interactions\cite{jaschke2017critical}.

\begin{figure}[tbp]
 \begin{center}
\includegraphics[scale=0.25]{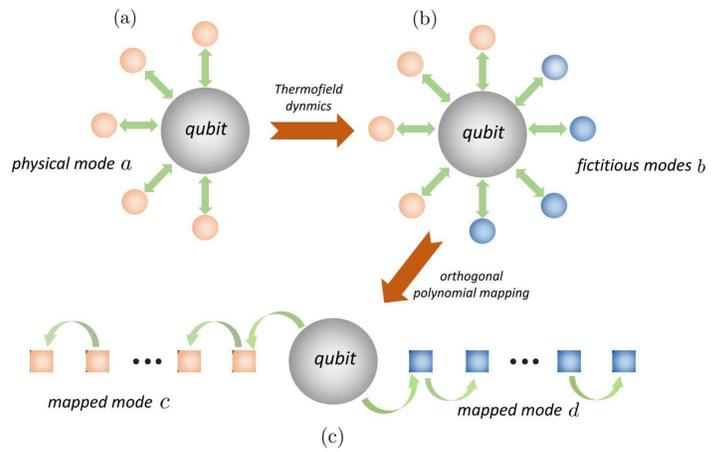}
\end{center}
\caption{(Color online) Diagrammatic description of our numerical method. (a) The original Hamiltonian of the open quantum system of a qubit coupling to its environment(boson mode $a$) (b) To ensure the efficiency of the MPS basis, the temperature effect is introduced by adding a group of fictitious bath modes(boson mode $b$) according to the TFD theory. (c) Both physical and fictitious bath modes are further mapped into a half-infinite chain of boson modes $c,d$  with only nearest interactions.}
\label{Fig0}
\end{figure}

In summary,  the methodology adopted in this paper can be described with the following steps(as depicted in Fig.(\ref{Fig0})): Firstly, according to the  TFD method, the original Hamiltonian  (\ref{SBH1}) is modified to (\ref{SBTFD}) by adding a fictitious environment to incorporate the finite temperature effect. Then, Hamiltonian (\ref{SBTFD}) is further mapped onto a one-dimensional chain representation ($\ref{FinalHami}$) with only nearest interactions. At last, the MPS-TDVP method is applied to simulate the evolution of Hamiltonian ($\ref{FinalHami}$).  Specially, $c_0^L =0 $ for zero temperature case, thus the Hamiltonian ($\ref{FinalHami}$) reduces to the semi-infinite chain structure which is  used in our previous study\cite{he2017zeno} of QZE in power-law spectral environment at zero temperature.

\subsection{Numerical examples}

\begin{figure}[tbp]
 \begin{center}
\includegraphics[scale=0.6]{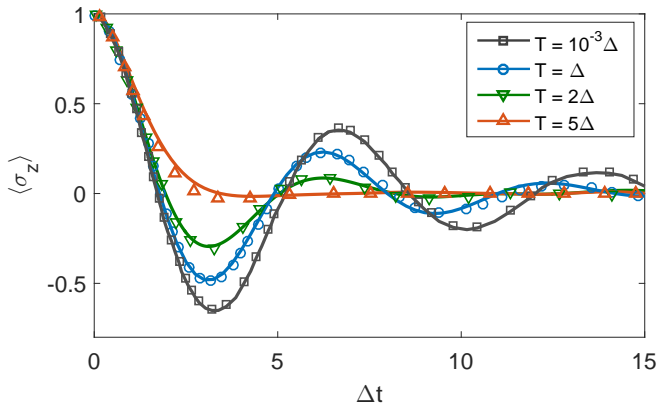}
\end{center}
\caption{(Color online) The time evolution of the qubit population inversion $\langle\sigma_z(t)\rangle$ in an Ohmic bath for different temperatures. Solid lines are results from MPS-TDVP-FTD method. The hollow marks are results of numerically exact HEOM method collected from Fig(3) in Ref\cite{yan2016stochastic}. Other parameters are $\eta = 0.4, \omega_c = 4\Delta$.  }
\label{Fig1}
\end{figure}

In order to show the validity of our numerical method, we compare our numerical results with the benchmark result obtained by HEOM\cite{tanimura1989time,tanimura2006stochastic}. The Hamiltonian of the qubit $ H_{\text{sys}}$ and the coupling operator $\hat{A}$ in (\ref{SBH1}) are chosen as:
\begin{align}
  H_{\text{sys}} = \frac{\Delta}{2}\sigma_x \quad\quad \hat{A} = {\sigma_z}
  \label{Hsys}
\end{align}
The spectral function of the bath is assumed to be an Ohmic form with the Debye regulation:
\begin{align}
 J(\omega) = \eta \omega \omega_c^2/(\omega_c^2 +\omega^2)
\end{align}
where $\eta$ is the coupling strength and $\omega_c$ is the cut-off frequency. We calculate the evolution of the expectation $\sigma_z(t) = \text{Tr}\left[\sigma_z\rho_{\text{sys}}\right]$  of the qubit and compare results of our numerical method with those obtained from HEOM. As depicted in Fig(\ref{Fig1}), it shows a good agreement in a large range of temperature. Different from HEOM, our numerical method is not limited to the form of the bath spectral function. Thus we will apply this reliable numerical method to simulate Zeno dynamics of a qubit in a 1/f noise bath.

\section{QZE and QAZE of an open quantum system in 1/f noise bath}

Now we turn to the QZE and QAZE in an open quantum system of a qubit interacting with a 1/f noise bath. This situation can be described by Hamiltonian (\ref{SBH1}) with $H_{\text{sys}}$ and $\hat{A}$ defined in (\ref{Hsys}). The spectral function of $1/f$ noise is specified as\cite{dutta1981low}:
\begin{align}
 J(\omega) = \frac{\alpha}{\omega}\Theta(\omega - \omega_0)\Theta(\omega_c-\omega)
\end{align}
Here $\Theta$ is the Heaviside step function: $\Theta(x)=1$ for $x>0$ and vanishes elsewhere. $\alpha $ is the coupling strength, $\omega_0, \omega_c$ are lower and upper cutoff frequencies respectively. Without loss of generality, we set $\omega_0 = 0.1\Delta, \omega_c =10\Delta$ in the rest of the paper. We focus on the initial state of a product state for both zero($\beta\rightarrow \infty$) and finite temperature cases:
\begin{align}
 \rho(0) = |e\rangle\langle e|\otimes \rho_B(\beta)\label{InitialState}
\end{align}
where $|e\rangle$ is the excited qubit state as $\sigma_z|e\rangle =|e\rangle$ and $\rho_B(\beta) = e^{-\beta H_{\text{env}}}/\text{Tr}\left(e^{-\beta H_{\text{env}}}\right)$ is the thermal equilibrium state of the environment.
%We first discuss the QZE and QAZE at zero temperature.

The  QZE(QAZE) can be described by the survival probability $P_\text{sur}(t=n\tau)$ which is defined as the probability of finding the initial state after $n$ successive measurements with equal time interval $\tau$. The measurement considered in this paper is assumed to be an ideal projection known as the wave-packet collapse postulation\cite{von1955mathematical} which states that the measured quantum system(qubit) is completely collapsed to its initial state after instantaneous measurement. Consequently, the survival probability under successive measurements can be written as:
\begin{align}
 P_{\text{sur}}(t=n\tau) = \text{Tr}\left[ P_{M}e^{-iH\tau}\rho(0,\beta)e^{iH\tau}P_{M} \right]^n = P^n_{\text{sur}}(\tau) \label{PsurWeak}
\end{align}
where $P_{M} = |e\rangle\langle e|$ is measurement projecting operator. In the short interval time limit $\tau \rightarrow 0$,  one can further write $P_{\text{sur}}(t=n\tau)$ in an exponentially decay form\cite{facchi2001quantum,maniscalco2006zeno}:
\begin{align}
 P_\text{sur}(n\tau) = \exp(-\Gamma(\tau)t)
\end{align}
where $\Gamma(\tau)$ is  an effective decay rate:
\begin{align}
 \Gamma(\tau) = -\frac{1}{\tau}\ln\left[P_\text{sur}(\tau)\right]\label{GammaLog}
\end{align}
This effective rate $\Gamma(\tau)$ is valid, only if the system-bath coupling strength is weak enough or the measurement is performed sufficiently fast.
Hence, the memory effect between measurements can be neglected. One can expect that this measurement induced non-markovian effect should have a significant influence on the QZE and QAZ in the 1/f noise bath due to its long time correlations.  The general definition of the survival probability  that incorporates such non-markovian effect can be modified as\cite{chaudhry2014zeno,zhou2017quantum}:
\begin{align}
 P_{\text{sur}}(t=n\tau) &= \text{Tr}\left[ \left(P_Me^{-iH\tau}\right)^n\rho(0,\beta)\left(e^{iH\tau}P_M\right)^n \right]\notag\\
 &= \prod_{i=1}^n \tilde{P}_{\text{sur}}(\tau,i)
 \label{ExactPsur}
\end{align}
where $\tilde{P}_{\text{sur}}(\tau,i)$ is the $i$th measurement survival probability whose initial state(renormalized) is obtained from $(i-1)$th measurement projection.
Although the decay of the modified survival probability (\ref{ExactPsur})may not follow an exponential form, the effective decay rate defined in Eq.(\ref{GammaLog}) can still be regarded as a quantity to describe the speed of survival probability's decay due to the monotonic relation between $\Gamma$ and $P_\text{sur}$.  However, $\Gamma$ now is dependent on both measurement interval $\tau$ and the number of measurements  $n=t/\tau$:
\begin{align}~\label{gamma}
 \Gamma(\tau,t=n\tau) = -\frac{1}{n\tau}\ln\left[ P_{\text{sur}}(t=n \tau)\right]
\end{align}

The effective decay rate $\Gamma(\tau,t=n\tau)$ is crucial quantities to characterize QZE and QAZ\cite{zheng2008quantum,cao2010dynamics},
and reduces to $\Gamma(\tau)$ at Eq.~(\ref{GammaLog}) by taking $n=1$.
However,  The direction application of $\Gamma$ at Eq.~(\ref{gamma}) may be inconvenient for numerical methods.
In this paper, we alternatively adopt a derivative of $\Gamma$, namely $\frac{\partial \Gamma}{\partial \tau}$, to classify QZE and QAZE: $\frac{\partial \Gamma}{\partial \tau}>0$ means that the system is more severely slowed-down by faster repeated measurements, indicating the occurrence of QZE; on the contrary, $\frac{\partial \Gamma}{\partial \tau}<0$ can be regarded as the characteristic of QAZE since the decay is accelerated by frequent measurements.
This definition retains the core physical picture of QZE and QAZE without calculating $\Gamma_0$, which has been commonly used in recent studies\cite{chaudhry2014zeno,wu2017quantum,he2017zeno}.
Throughout this paper, we use this new definition of  QZE and QAZE.

\subsection{zero temperature case}

\begin{figure}[tbp]
 \begin{center}
\includegraphics[scale=0.65]{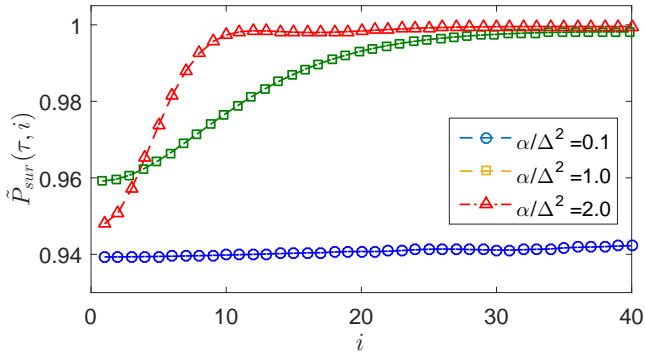}
\end{center}
\caption{(Color online) The evolution of the qubit's survival probability for each measurement number with $\Delta\tau = 0.5$ at zero temperature. For the weak coupling strength $\alpha/\Delta^2 =0.1$(solid line), $P_\text{sur}$ almost keeps unchanged, which indicates the validity of Eq.~(\ref{PsurWeak}). For the strong coupling strength $\alpha/\Delta^2 = 1.0,2.0$(dashed and dashed-dot line) $\tilde{P}_\text{sur}(\tau,i)$ gradually evolves towards a steady value, showing that successive measurements drive the qubit with its environment into a new dynamical equilibrium state.}
\label{Fig2}
\end{figure}

We firstly analyze the QZE at zero temperature,  as shown in Fig(\ref{Fig2}).
We calculate the $\tilde{P}_{\text{sur}}(\tau,i)$ at each measurement with $\Delta\tau = 0.5$. For the weak coupling strength($\alpha/\Delta^2 = 0.1$), the time evolutions of $\tilde{P}_{\text{sur}}(\tau,i)$ stay unchanged indicating that the system as well as its environment collapse to the same initial state ($\ref{InitialState}$) after each instantaneous measurement. Interestingly, with the increase of the coupling strength($\alpha/\Delta^2 =1.0\sim2.0$ ), $\tilde{P}_{\text{sur}}(\tau,i)$  exhibit a relaxation process: the measurement-induced non-Markovian effect breaks the periodicity of the evolution observed in the weak coupling regime and drives the system along with the environment to a new dynamical-equilibrium state in the long time scale. We can observe a suppression of the decay of $\tilde{P}_{\text{sur}}(\tau,i)$ with the increase of measurement number $i$. An intuitive explanation can be given as the return of the energy back-flow from the environment which repopulates the qubit to the initial excited state. {With the consideration of the measurement-induced non-Markovinity, the measurement projections partially retain the energy of the bath stored in the pervious free evolution(compared to the thermal equilibrium state at zero temperature). In the next evolution-measurement cycle, this excited bath pumps the energy back to the system, assists to increase the population of the excited(initial) state of the qubit.
This repopulation assisted  by measurements has also been studied in the framework of dynamic interpretation of QZE in ref\cite{ai2013quantum}, where measurements are realized by coherently pumping to a third level and spontaneously decay back to the the lower state.
Hence, we can infer that with the increase of the system-bath coupling strength as well as the measurement interval $\tau$, this effect of the energy back-flow will more seriously repopulate the qubit, leading to a greater suppression of the decay of the survival probability $P^{(i)}_{\text{sur}}$.}

\begin{figure}[tbp]
 \begin{center}
\includegraphics[scale=0.5]{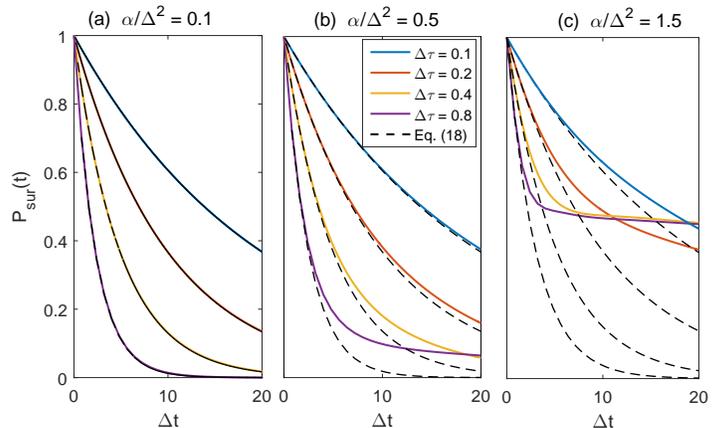}
\end{center}
\caption{(Color online) The survival probability $P_{\text{sur}}$ with different measurement intervals from weak to strong coupling strengths. For the weak coupling strength($\alpha/\Delta^2 = 0.1$), the exponential decay rate is still valid and a Zeno effect is predicted. However, for $\alpha/\Delta^2 = 0.5$, the decay of $P_{\text{sur}}$  gradually deviates from the exponential form with increase of measurement interval $\tau$. Such deviation is more apparent in the strong coupling regime($\alpha/\Delta^2 =1.5$) where the decay of $P_{\text{sur}}$ demonstrates an obvious suppression for $\Delta\tau =0.4,0.8$ compared to the relatively shorter measurement intervals $\Delta\tau = 0.1,0.2$ after several measurements. This demonstrates a crossover from Zeno to anti-Zeno along time  in the strong coupling regime.
}
\label{Fig3}
\end{figure}

{To show the validity of this inference}, we analyze the influence of the measurement-induced disturbance on the survival probability.
In Fig(\ref{Fig3}), we calculate $P_\text{sur}(t)$ for different measurement intervals $\tau$ from weak to strong coupling strengths.  For the weak coupling strength ($\alpha/\Delta^2 = 0.1$), the memory effect of the bath between nearest measurements can be neglected and the bath is recovered to the original thermal equilibrium state after each projecting. Thus $P_\text{sur}$ demonstrates an exponential decay as predicted by previous studies.  However, when increasing $\alpha$ to $\alpha/\Delta^2 = 0.5$, the dynamical behaviours of the decay of $P_{\text{sur}}$ gradually deviate from an exponential form with increase of the measurement interval $\tau$. A suppression on the decay of $P_\text{sur}$ is observed compared to results  without considering the measurement-induced non-Markovian effect(dashed line). This suppression effect are more serious when further increase the coupling strength to $\alpha/\Delta^2 = 1.0$ even for a short measurement interval $\Delta\tau >0.2$. Interestingly, we can observe an intersection of $P_{\text{sur}}$ between $\Delta\tau = 0.2$ and $\Delta\tau = 0.4,0.8$ in (\ref{Fig3}.c).
As a consequence, we find a novel crossover from pure QZE in the short time scale(or small measurement number $n$) to a QZE-QAZE transition in the long time scale(or large measurement number). It is entirely induced by the non-Markovian feedback of the environment through repetitive measurements.

\begin{figure}[tbp]
 \begin{center}
\includegraphics[scale=0.6]{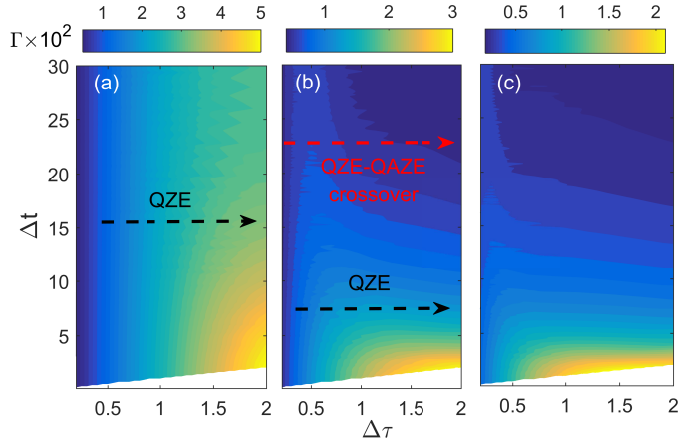}
\end{center}
\caption{The effective decay rate  $\Gamma(\tau,t)$ for different coupling strength (a)$\alpha/\Delta^2 = 0.1$ ,(b)$\alpha/\Delta^2 = 0.5$, (c)$\alpha/\Delta^2 = 1.0$.  (a): $\Gamma(\tau,t)$  monotonically increases with $\tau$ for the whole range of $t$, demonstrating an pure QZE in the weak coupling regime for both short  and long time scale. (b): $\Gamma(\tau,t)$ suffer an obvious suppression with the increase of $t$, especially for relatively large measurement intervals($\Delta\tau>1$). This leads to an obvious division along $t$ where pure QZE is observed in the short time scale($\Delta t<10$) while QZE-QAZE crossover in the long time scale($\Delta t>10$). (c): The suppression of the decay rate $\Gamma(\tau,t)$ is more remarkable in the long time scale, even for short measurement intervals $\Delta\tau\approx 0.15$.
%Note all data of $\Gamma(\tau,t)$  are well defined only for $t=n\tau$, we fit rest of data to plot the contour for clarity.
\label{Fig4}}
\end{figure}

To show this novel crossover clearly, we calculate the time-dependent effective decay rate $\Gamma(\tau,t)$ for different coupling strength and measurement intervals, as shown in  Fig(\ref{Fig4}). For the weak coupling strength $\alpha/\Delta^2 = 0.1$,  $\Gamma(\tau,t)$ monotonically increase with the increase of the measurement interval $\tau$, demonstrating a QZE according to the definition. Nevertheless, $\Gamma(\tau,t)$ hardly vary with $t$, indicating that the bath memory effect between measurements is weak and the evolution has little dependence on the measurement number.  By increasing the coupling strength to $\alpha/\Delta^2 =0.5$(Fig(\ref{Fig4}.b)),  an obvious suppression of the $\Gamma(\tau,t)$  with the increase of $t$ can be observed, especially for relatively large measurement intervals($\Delta\tau >1.0$). It is astonishing that we find a novel crossover of pure-QZE to QZE-QAZE behaviour along $t$(or the measurement number):  the effective decay rate $\Gamma(\tau,t)$  monotonically increases with $\tau$  showing a pure QZE in the short time scale($\Delta t<8)$. However, in the long time scale($\Delta t>10$), the effective decay rate $\Gamma(\tau,t)$ first increases and then decreases with the increase of the measurement interval. As already shown in Fig(\ref{Fig3}),  the alteration of the initial bath state caused by  each measurement projection leads to a relaxation process in which the decay of the survival probability is gradually suppressed. It is this long time accumulation of the suppression which results in this time-scale-divided crossover. By further increasing the coupling strength to $\alpha/\Delta^2 =1.0$(Fig(\ref{Fig4}.c)), a more remarkable suppression to $\Gamma(\tau,t)$ can be observed even for short measurement intervals $\Delta\tau \approx 0.4$.

It should be noted that this repopulation phenomenon is different from the case of the bath with power-law spectral function(Ohmic and sub-Ohmic bath) in our previous study\cite{he2017zeno}. In that case, we observed an increase of the decay rate for $P_{\text{sur}}$ due to the similar consideration of the measurements disturbance to the environment.  This suggest that the role of this measurement-induced non-Markovian effect on the survival probability $P_{\text{sur}}$ is sensitive to the bath spectral function and has a significant connection to the multi-photon excitation process due to the strong coupling strength between TLS and its environment. This is beyond the scope of this paper and will be studied in the future.

\subsection{finite temperature case}
\begin{figure}[tbp]
\begin{center}
\includegraphics[scale=0.45]{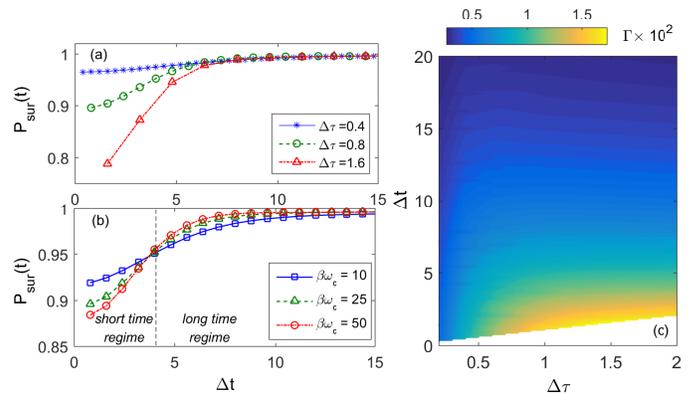}
\end{center}
\caption{The influence of the temperature on the survival probability $P_\text{sur}(t)$ and effective decay rate $\Gamma(\tau,t): $(a): The evolution of  $P_\text{sur}(t)$ for different measurement intervals $\Delta\tau = 0.4,0.8,1.6$(star,circle,triangle) at $\beta\omega_c = 25$. Similar to the case of zero temperature, successive measurements drive $P_{\text{sur}}$ to a constant, indicating the formation of the dynamical steady state. (b) The evolution of  $P_\text{sur}(t)$ at different temperature $\beta\omega_c = 10,25,50$(square,triangle,circle)  for $\Delta\tau =0.8$. The decays of $P_\text{sur}(t)$ are increasingly suppressed with the increase of the temperature in the short time regime while enhanced in the long time regime.  (c): Effective decay rate $\Gamma(\tau,t)$ for the coupling strength $\alpha/\Delta^2 =0.5$ at the finite temperature $\beta\omega_{{cu}} =15$. The crossover of the QZE-AQZE occurred at zero temperature in the long time regime (in Fig(\ref{Fig4}.b)) can hardly be observed in this case. Note to make the figure clear, only data of $P_\text{sur}(t)$ just before measurement projection are presented in Fig (\ref{Fig5}.(a)).
}
\label{Fig5}
\end{figure}

In this section, we extend our study to the finite temperature case.  In order to understand the role of the thermal fluctuation during repeated measurements, we first calculate $P_\text{sur}(t)$ for different temperature, as shown in Fig ($\ref{Fig5}$.a). When the measurement interval is relatively short($\Delta \tau =0.4$), the finite temperature effect has almost no influence on dynamics of $P_\text{sur}(t)$. By increasing the interval to $\Delta\tau = 0.8$, opposite effect of the finite temperature depending on the time-scale $t$  can be observed: on the one hand, $P_\text{sur}(t)$ is suppressed with the increase of the temperature in the short time regime before $\Delta t<4$; on the other hand, higher temperature leads to a faster decay of  $P_\text{sur}(t)$ for $\Delta t>4$ until the dynamical equilibrium state is reached.
The typical time that dividing the suppression or enhancement of the $P_\text{sur}(t)$ decay is advanced by increasing the measurement interval, as for $\Delta\tau = 1.6$ the suppression of the decay only lasts for first two rounds of measurements($\Delta t\approx 3$).
The suppression of $P_\text{sur}(t)$ caused by finite temperature can be qualitatively described by the noninteracting blip approximation (NIBA)\cite{dekker1987noninteracting}.
Including the Born approximation and Silbey-Harris transformation\cite{silbey1984variational}, NIBA can be realized based on the second-order perturbation master equation.
Then, the evolution of $\langle\sigma_z(t)\rangle$(which is proportional to $P_\text{sur}(t)$ ) in a thermal bath can be written as:
\begin{align}
  \frac{d\langle\sigma_z(t)\rangle}{dt} +\int_0^t f(t-t')\langle\sigma_z(t')\rangle dt' =0
\end{align}
with the kernel function $f(\tau)$ defined as:
\begin{align}
 &f(\tau) =\Delta^2\cos\left[Q_1(\tau)\right]\exp[-Q_2(\tau)]\notag\\
 & Q_1(\tau) \equiv \int_0^\infty \sin(\omega \tau)J(\omega) d\omega/\omega^2\notag\\
 & Q_2(\tau) \equiv \int_0^\infty[1-\cos(\omega\tau)]\coth(\frac{1}{2}\beta\omega)J(\omega)d\omega/\omega^2
\end{align}
In the short time regime, the environment is still close to the thermal equilibrium state and equations above can be regard as valid. By increasing the temperature, $Q_2(\tau)$ is increased while  $f(\tau)$ is decreased for all $\tau$. This leads to the suppression of the non-Markovian effect represented by the integration of the kernel function. Consequently, the energy driven from the bath  as well as the decay $P_{\text{sur}}(t)$ is obstructed due to thermal fluctuations. In the long time regime during which the qubit and the bath relax to the dynamical steady state, it is the repopulation induced by the backflow of the  energy from the environment to the system that dominates the dynamics. The suppression of the the energy driven from the system in the short time regime now again affects this repopulation process: a less energy-absorbed environment results in less capability to repopulate the qubit. Both the short time and long time effects of the thermal fluctuations reduce the difference between pure QZE in the short scale and  QZE-QAZE transition in the long time scale . In Fig(\ref{Fig5}.b), we show the   effective decay rate $\Gamma(\tau,t)$ for the temperature of $\beta\omega_c =15$ and $\alpha/\Delta^2 =0.5$. Compared with Fig(\ref{Fig4}.b), one can find that $\Gamma(\tau,t)$ is obviously suppressed for $\Delta \tau >1$ where the maximal value decreases from $\Gamma = 3.0$ at zero temperature to $\Gamma = 1.5$. On the other hand, the disappearance due to the temperature induced suppression of the survival probability in short time scale. {In conclusion, the finite temperature induces two opposite effects to the survival probability according to the time scale:the thermal fluctuations raise an additional suppression of decay of the survival probability in the short time scale while  enhances the decay rate in long time scale. This finally make the transition of QZE to the crossover of QZE-QAZE along measurement number more difficult to observe.}

\section{Conclusion}

In summary, we study the QZE and QAZE of a qubit interacting with an environment of 1/f noise  with a numerically exact method based on the combination of  MPS and TFD.   By considering the non-Markovian effect induced by measurements which alters the initial state of the bath after each measurement projection, the evolution demonstrates a relaxation process that drives both qubit and the bath to an non-equilibrium dynamical steady state. The effective decay rate $\Gamma(\tau,t=n\tau)$ under this situation is dependent on both measurement  interval $\tau$ and evolution time $t$(or number of measurements$n$). At zero temperature, we observe a novel transition from a pure QZE in the short time scale(small $n$) to a crossover of QZE-QAZE in the long time scale(large $n$)  with the increase of system-bath coupling strength. This is due to the energy back-flow from the bath which repopulates the qubit to the initial excited state and suppresses the decay of the survival probability. Moreover, we generalized our study to the finite temperature case. We find that the thermal fluctuations raise an additional suppression of decay of the survival probability in the short time scale while enhances the decay rate in long time scale.

\section{Acknowledgement}
Chen Wang is supported by the National Natural Science Foundation of China under Grant Nos. 11704093 and 11547124.
Qing-hu Chen is supported by the National Natural Science Foundation of China under Grant Nos.11674285 and 11474256.

\newpage

\bibliographystyle{unsrt}%
\bibliography{Zeno1OF}

\end{document}